# Short pulse close to round-trip time generated by cavityless high gain Nd:GdVO$_4$ bounce geometry

**RUI GUO,**[1,†] **MINGMING NIE,**[1,2,†] **QIANG LIU,**[1,2] **MALI GONG**[1,3,*]

[1] *State Key Laboratory of Precision Measurement and Instruments, Tsinghua University, Beijing 100084*
[2] *Key Laboratory of Photon Measurement and Control Technology, Ministry of Education, Tsinghua University, Beijing 100084, China*
[3] *State Key Laboratory of Tribology, Tsinghua University, Beijing 100084*
*\* Corresponding author: gongml@mail.tsinghua.edu.cn*

**Abstract:** In this paper, laser pulses with pulse-widths approach to the round-trip time are generated by utilizing a cavityless high gain Nd:GdVO$_4$ bounce geometry. By adopting an EOQ (electro-optics Q-switch), pulse-widths of 1.36 ns, 1.82 ns, and 2.39 ns are achieved at three effective cavity lengths respectively. All these pulse-widths are close to the round-trip time of corresponding effective cavity lengths. Moreover, the output power reaches watt-level and the repetition rate is kHz-level, meanwhile the M$^2$ factor is less than 1.3. Spectrally, the laser has a continuous spectrum with 10 dB linewidth of 0.2 nm.

## 1. Introduction

Short nanosecond laser pulses are usually desirable in many industrial and medical applications[1-5]. Q-switch technique is capable of producing laser pulses with durations from sub-nanosecond to several tens of nanoseconds. By utilizing Q-switch devices, cavity losses are periodically modulated, and laser pulses are built up in the cavity and transmit through the output coupler. Generally, the building up of Q-switch pulses requires several to several tens of round-trips in the cavity. In order to achieve short pulses, the cavity lengths are normally reduced, such as microchip lasers with passive Q-switch[2, 3]. However, short cavity lengths will bring the small fundamental mode size, leading to the restriction on the output power.

Another approach to minimize the pulse-width is increasing the gain. With higher gain, the number of round-trips necessary to generate the pulse can be reduced[1, 4, 5]. Besides, the pulse-width can also be minimized by optimizing the transmittance of the output coupler at a given gain level. In 1989, John J. Degnan proposed the theory of optimally coupled Q-switch laser[6]. Zayhowski et al. also investigated the optimized output coupling for minimizing pulse-width [7, 8]. It reveals that the optimized output coupling is increased as with the increase of gain level. This implies an output coupler with 100% transmittance, namely a cavityless configuration, will satisfy the optimal condition of minimizing pulse-width under an extremely high gain.

In this paper, a cavityless high gain bounce geometry is adopted to produce short laser pulses for the first time, to the best of our knowledge. With the high gain provided and the cavityless configuration applied, experiments of three effective cavity lengths are implemented with an EOQ (electro-optics Q-switch) device. Minimum pulse-widths of 1.36 ns, 1.82 ns and 2.39 ns are achieved respectively, which are all close to the corresponding round-trip time. Moreover, the pulse-energy, beam quality and spectral characteristics are also investigated in this work.

## 2. Experimental Setup

The cavityless configuration here is similar to that in Smith's studies[9, 10]. As presented in Fig. 1, the cavityless laser adopts a high gain bounce geometry without using an output coupler.

The laser output is investigated for several performances, including output power, spatial beam quality, pulse characteristics and spectral characteristics. Divided by a beam splitter (BS), one path with most of the light power enters into the power meter. Another path of the laser is used to measure the spatial beam quality and the spectral characteristics. The pulse characteristics are measured by detecting the diffused light scattered from the power meter.

## 3. Results and discussions

### 3.1 CW output power and beam quality

We firstly realize a CW mode laser output without Q-switch devices working. Here we use $L_{eff}$ to represent the total optical path between $M_0$ and the exiting surface of the crystal. Three experiments under $L_{eff} = 170, 240$ and $310$ mm are implemented. At the pump power of 30 W, the output power are 4.5 W, 4.0 W and 3.6 W for three $L_{eff}$ respectively, corresponding to the optical-optical efficiency of 15%, 13% and 12%. The output power against pump power for $L_{eff}$=170 mm is shown in Fig. 2.

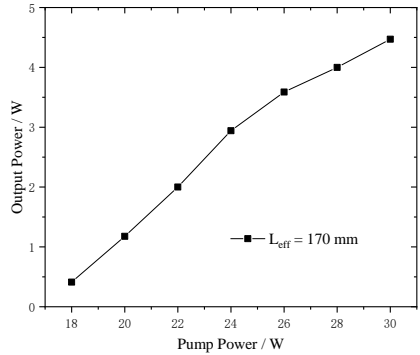

Fig. 2. Output power against pump power

Using a $M^2$ beam analyzer (Spiricon $M^2$-200), the $M^2$ factor is measured for three $L_{eff}$. The results of $M_x^2$ <1.3 and $M_y^2$ <1.2 can be realized for all three $L_{eff}$ at their highest output power, which is shown in Fig. 3. The inset shows the spatial profile of far-field spot.

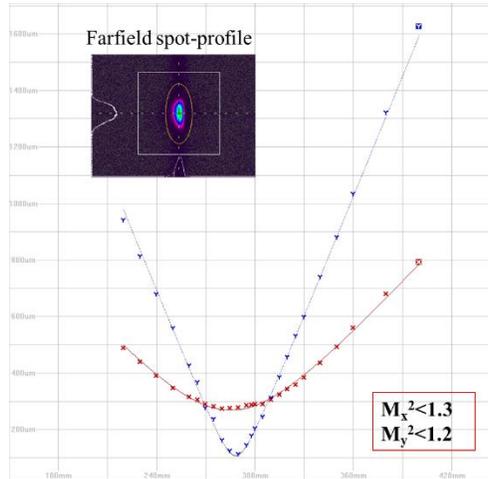

Fig. 3 Beam quality of the cavityless laser

*3.2 Pulse characteristics*

The pulsed cavityless laser is carried out at the same three $L_{eff}$ values. During the pulsed experiment, the pulse profile is measured by an InGaAs detector (bandwith of 25 GHz) and an oscilloscope (Tektronix MSO 73304 DX, sample rate of 100 GS/s and bandwidth of 33 GHz). The pulse profile is also measured at a bandwidth of 1 GHz in order to characterize the envelope of the waveform. Fig. 4 (a) presents the measured FWHM (full width at half maximum) of laser pulses at the pump power of 30 W for the repetition rate ranging between 1 and 20 kHz at three $L_{eff}$. It is found that the pulse-width rises monotonously for all three $L_{eff}$. At the repetition rate of 1 kHz, the pulse-widths of $L_{eff} = 170, 240$ and $310$ mm reach 1.36 ns, 1.82 ns and 2.39 ns, respectively. The corresponding pulse profiles for bandwidth of 1 GHz and 25 GHz are both presented in the Fig. 4 (b). It is worth noting that the round-trip time of the corresponding $L_{eff}$ is 1.13 ns, 1.60 ns, and 2.07 ns, respectively. This indicates that pulse-widths in the pulsed cavityless Nd:GdVO$_4$ bounce laser at 1 kHz repetition rate has reached 1.20, 1.13 and 1.15 times round-trip time of corresponding $L_{eff}$. As with the increase of repetition rate, the pulse-width broadens more rapidly for longer $L_{eff}$. This is due to the diffraction loss becomes larger for the longer $L_{eff}$. For the $L_{eff} = 310$ mm, the pulse-width broadens to 6.76 ns at 20 kHz (3.3 times the corresponding round-trip time), while for the $L_{eff} = 170$ mm, the pulse-width only broadens to 1.90 ns (1.7 times the corresponding round-trip time). These short pulse-widths comparable to the round-trip time have approached to the limitation of pulse-widths in Q-switched lasers, because in ordinary Q-switched lasers, at least several round-trips are required to build up the pulses and sweep out the gain in the crystal. According to Zayhowski's theory, the optimized output coupling for realizing the minimum pulse-width in a Q-switched laser is[7]:

$$\gamma_0 = 0.33 N_0 \sigma l_r - \gamma_p \qquad (1)$$

where $\gamma_0$ is the parameter of output coupling. $N_0$ represents the initial inversion density, and $\sigma$ is the stimulated emission cross-section of the laser medium. $l_r$ is the round-trip intracavity beam path length. $\gamma_p$ is the parameter of round-trip parasitic loss. The transmission of output coupler $T_0$ can be derived by $\gamma_0$, as shown below:

$$T_0 = 1 - e^{-\gamma_0}. \qquad (2)$$

Eq. (1-2) indicates that the optimal output coupling to achieve short Q-switched pulses will approach to 100% if there is sufficiently high gain. In our experiment, the single-pass small signal gain is measured to be $2 \times 10^4$. If we treated the term $e^{N_0 \sigma l_r/2}$ as single-pass small signal gain, $T_0$ should be $\approx 0.997$. Hence, the optimal output coupling to achieve minimum pulse-width is nearly satisfied under the cavityless configuration. In this way, short pulse-widths close to round-trip time are achieved.

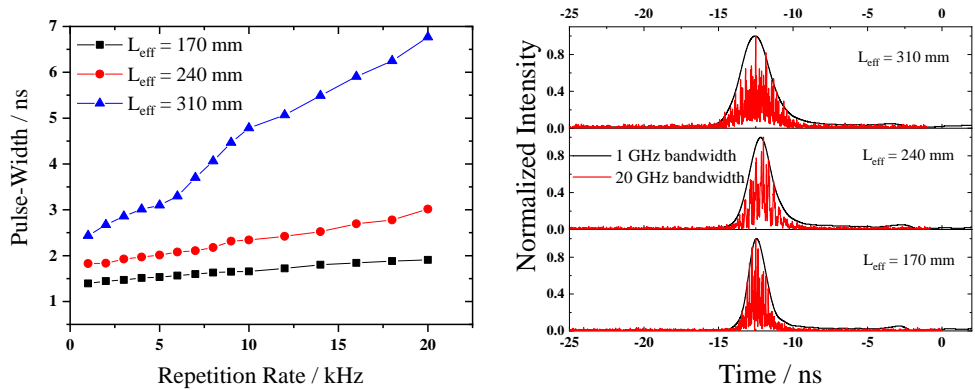

Fig. 4. Pulse characteristics of cavityless laser under three different $L_{eff}$: (a) Pulse-width against repetition rate, (b) Pulse profile at 1kHz

It is well-known that cavity dumping technique is capable of producing pulse-width that approach to $t_r$, but drivers with corresponding short rise/fall time are needed [12-14]. For the shortest $L_{eff}$=170 mm, a driver with switch time of ~1.36 ns is needed in cavity dumping method. However, in our experiment, the switch time of the EOQ driver is about 6 ns, which still results in the pulse-width of 1.36 ns. This indicates that a high gain cavityless configuration can obtain short pulses with sufficient tolerance to the speed of the Q-switch driver.

### 3.3 Pulse energy

The average output power (at the pump power of 30 W) that related to repetition rate is illustrated in Fig. 5 (a). With the increase of repetition rates from 1 kHz to 20 kHz, the average power increases from 1.11 , 0.96 and 0.47 W to 3.15, 2.76, and 2.29 W for $L_{eff}$=170, 250, and 340mm, respectively. Combined with the corresponding CW mode power, the Q-switched efficiency is 70%, 69%, and 63.6% at 20 kHz, accordingly. The extinction power corresponding to $L_{eff} = 170, 240$ and 310 mm is 0.6 W, 0.4 W and 0.2 W, respectively. Subtracting the extinction power, the pulse energy of a single pulse can be calculated, as illustrated in Fig. 5 (b). For the shortest $L_{eff}$, the pulse energy reaches 600 μJ at 1 kHz, thus the maximum peak power exceeds 440 kW. The average output power is above 2 W at 10 kHz, while the maintaining the pulse-widths of <1.5 times the round-trip time. These results demonstrate a high gain Nd:GdVO$_4$ bounce geometry with cavityless configuration is capable of obtaining wat-level laser output with pulse-widths close to round-trip time. During the varying of the repetition rate, the beam quality is monitored, and the M$^2$ factor stays lower than 1.3 all through on two directions.

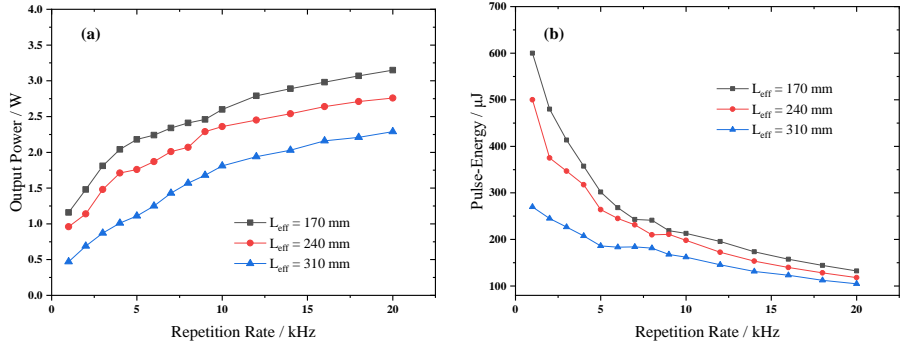

Fig. 5 Pulse energy characteristics of cavityless laser under three different $L_{eff}$ (a) output power against repetition rate, (b) pulse energy against repetition rate.

### 3.4 Spectral characteristics

The spectra of the cavityless laser output are monitored by an optical spectrum analyzer (Agilent 86142B, minimum resolution of 0.06nm) for CW mode at the pump power of 30 W. The 10 dB spectral linewidth of the output is 0.2 nm, as shown in Fig. 6.

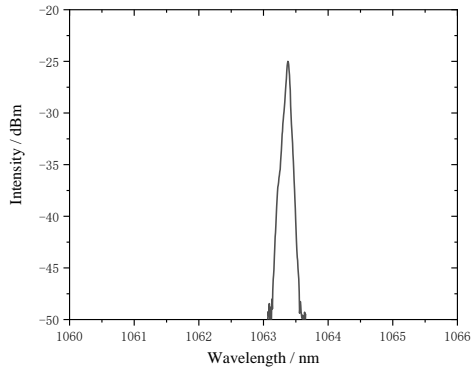

Fig. 6. Spectrum of the cavityless laser

As the resolution of 0.06nm cannot offer more specific spectral information, the spectral content is further investigated by a scanning Fabry-Perot (FP) interferometer with the free spectral range (FSR) of 3 GHz. After divided by BS, a small portion of power about 30mW enters into FP interferometer. When the cavityless laser output works at CW mode (still at a pump power of 30 W), the scanning output is the continuous spectrum accompanied by several disordered peaks: both the amplitudes and the spacings of these peaks are random, as shown in Single Frame 1-4 of Fig. 7 (a). If multiple frames are averaged (64 frames averaged in Fig. 7(a)), the scanning output becomes flatter. No steady longitudinal modes or longitudinal intervals are discovered. To investigate the spectral characteristics obtained here, we implement a comparative experiment. When an output coupler (T = 80%) is inserted to make up a cavity, the output power increases to 6.2 W. This indicates the oscillation is formed. At this moment, 4 peaks with fixed interval are observed in the scanning output, which implies at least 4 steady longitudinal modes are generated. The measured longitudinal mode intervals are consistent with the cavity length of the oscillator. This phenomenon proves the previous observed random emerging peaks are not caused by multiple longitudinal modes. The random emerging peaks are averaged to be continuous in the range of 3 GHz, which implies the spectra characteristics approach to that of ASE in time-average observation.

The spectra of pulsed operating cavityless laser output are also analyzed in FP interferometer. When the cavityless laser is working at the pulsed state (100 Hz, 200 Hz and 500 Hz), the scanning output of FP interferometer has equal repetition rate with that of the cavityless laser, as illustrated in Fig. 7(b). The result is similar to that in the frequency-shifted feedback laser [15] and the narrow-linewidth Nd:YAG ASE source [16], both of which own a continuous spectrum. Considering the results in Fig. 7 (a) and (b), it can be concluded that the output of the cavityless configuration in our experiment has a continuous spectrum rather than steady longitudinal structures. This spectral characteristic might be helpful in applications such as nonlinear interactions[17, 18].

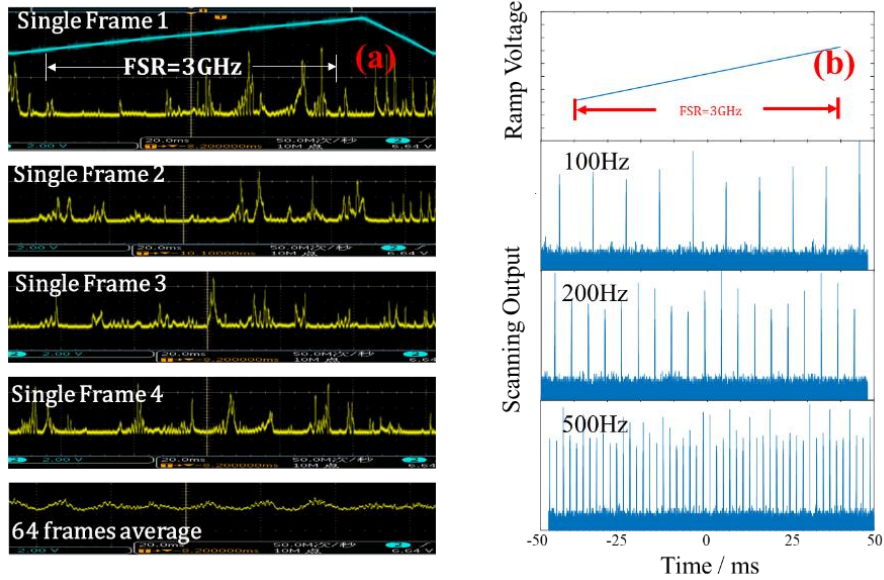

Fig. 7 FP scanning output of the cavityless laser (a) CW mode, (b) pulsed mode

## 4. Conclusion

In summary, a watt-level, kHz-level laser output is obtained with pulse-widths close to the round-trip time. The beam quality factor is near the diffraction limit ($M^2 < 1.3$). The spectrum is continuous and has a 10 dB linewidth of 0.2 nm. The minimum pulse-widths reach 1.36 ns, 1.82 ns, and 2.39 ns at three cavity lengths respectively, which are all close to the corresponding round-trip time. The short pulse-widths close to round-trip time have approached to the limitation of the pulse-widths in Q-switched lasers. This is attributed to the setup combining the cavityless configuration and the high gain Nd:GdVO$_4$ bounce geometry. This study supplies a convenient method to produce short nanosecond pulses with watt-level power and kHz repetition rate, which is of great potential in many industrial and civilian applications.

## Funding

This research is supported by National Key Research and Development Program of China (2017YFB1104500).

† These authors contributed equally to this work